\documentclass{elsart}
\usepackage{graphicx}
\usepackage{amssymb}
\begin{document}
\begin{frontmatter}
\title{Cascade Training Technique for Particle Identification}
\author{Yong Liu\corauthref{yong}},
\corauth[yong]{Corresponding author.}
\ead{yongliu@fnal.gov}
\author{Ion Stancu}
\address{Department of Physics and Astronomy\\
         University of Alabama, Tuscaloosa, AL 35487}
%
\begin{abstract}
The cascade training technique which was developed during our work on the
MiniBooNE particle identification has been found to be a very efficient way to
improve the selection performance, especially when very low background
contamination levels are desired.
The detailed description of this technique is presented here based on the
MiniBooNE detector Monte Carlo simulations, using both artifical neural
networks and boosted decision trees as examples.      
\end{abstract}
\end{frontmatter}
%
\section{Introduction}

Particle identification (PID) is the procedure of selecting signal events while
rejecting background events, and is a key step in the data analysis for
essentially all particle physics experiments.
For some experiments, the search for a possibly very small signal within
a large amount of data, such as MiniBooNE, requires an extremely powerful
separation of signal from background events.

MiniBooNE~\cite{boone} is a short baseline accelerator neutrino experiment
currently running at the Fermi National Accelerator Laboratory.
Its primary goal is to definitely confirm or rule out the potential
$\bar\nu_\mu \to \bar\nu_e$ oscillation signal claimed by the LSND
experiment~\cite{lsnd}, by looking in the CP-conjugate channel for the $\nu_e$
appearance in a $\nu_\mu$ beam.
The $\nu_e$ appearance is identified by the presence of an isolated electron
from the charged-current $\nu_eC \to e^-X$ reaction on the carbon atoms of the
mineral oil active medium.
In addition to the intrinsic $\nu_e$ contamination in the beam, the main
backgrounds to the oscillations analysis come from misidentified muons and
neutral pions from the $\nu_\mu$ quasielastic scattering and neutral-current
$\pi^0$ production, respectively.
In order to reach the design sensitivity, the MiniBooNE PID algorithms must
achieve an electron selection efficiency of better than 50\%, for an overall
background contamination level of approximately 1\%.

The crucial points for any PID-based analysis are both the input variables and
the underlying algorithm.
Each variable must (obviously) have a relatively good separation between
signal and background events, while in addition it must yield a good agreement
between data and Monte Carlo (MC) in both distributions {\em and} correlations.
Once the set of PID variables has been identified, the next step is to chose
the PID algorithm.
In the case in which the signal event statistics are large enough and some few
variables already show a very clear separation between signal and background,
the simple, straight cuts method may be preferred.  
Otherwise, the linear or nonlinear combination techniques of the underlying
variables, such as Fisher discriminants~\cite{fisher}, artificial neural
networks (ANN)~\cite{ann-1}, or boosted decision trees (BDT)~\cite{boo-1}
should be applied.
While ANNs have been relatively widely accepted and successfully used in high
energy experimental data analysis in the past decades~\cite{ann-2}, BDTs as
a novel and powerful classification technique have been first introduced to the
particle physics community within the MiniBooNE collaboration only
recently~\cite{boo-2}.
Since then it has also been applied to the radiative leptonic decay
identification of B-mesons in the BaBar experiment~\cite{babar}, as well as to
supersymmetry searches at the LHC~\cite{conrad}.

Generally, given a particular set of PID variables and a particular selection
algorithm, the maximum PID performance is essentially fixed, up to relatively
small variations induced by adjusting some internal settings in the PID code
itself, e.g., the learning rate, number of hidden nodes, etc., in the case of
neural networks, or the number of leaves, minimum number of events in a leaf,
etc., in the case of decision trees.  
In addition to the PID variables and the algorithm, the training event sets
also play a crucial role in the PID performance.
Therefore, a careful selection of the training event sample may significantly
improve the overall PID performance, as developed within the cascade training
technique (CTT)~\cite{btn-178}. 

In this paper, both ANNs and BDTs are taken as examples to describe the cascade
training procedure, as based on the MiniBooNE detector MC simulations.
In Section~\ref{sec:vars} we describe briefly the MiniBooNE PID variables used
in this study, as well as a systematic procedure for the variable selection.
Section~\ref{sec:ctt} describes the cascade training technique, while the
results are discussed in Section~\ref{sec:results}.
Our conclusions are summarized in Section~\ref{sec:conclusions}.   
%
\section{PID variable construction and selection}\label{sec:vars}
%
The variable construction and selection is naturally a first concern for an
efficient PID.
The difference in the information content between signal and background events
based on which the separation is made has to be extracted from the variables
via some classification algorithm, such as Fisher discriminants, neural
networks, decision trees, etc.
The variable set used for different experiments and different analysis goals
will naturally differ from each other.
However, as already mentioned before, some fundamental requirements have to be
satisfied, namely:
(i)  the variable distributions must show some separation between signal 
     and background events, and
(ii) the variable distributions and their correlations must show good data/MC
     agreement.
The first requirement is directly connected with the maximum efficiency of the
PID, while the second one guarantees the relability of PID output.  

The MiniBooNE detector is a spherical steel tank of 610-cm radius, filled with
800 metric tons of ultra-pure mineral oil.  
An optical barrier divides the detector into an inner region of 575-cm radius,
viewed by 1280 photomultiplier tubes (PMT), while the 35-cm-thick outer volume,
viewed by 240 PMTs, serves as an active veto shield.
Neutrino interactions in the mineral oil are detected via both Cherenkov and
scintillation light with a ratio of 3:1 for highly relativistic particles.
The particle identification is essentially based on the time and charge
distribution recorded at the PMTs. 

Every event in the detector is subject to three different maximum-likelihood
reconstructions, assuming that the underlying event was an electron, a muon, or
a $\pi^0$.
Ideally, just the $e/\mu$ and $e/\pi$ likelihood ratios should be enough to
achieve a powerful separation of the electron signal from the muon and neutral
pion backgrounds.
However, this is not the case.
Therefore, in addition to these two variables, a large set of variables is
defined, based on the corrected time at the PMTs and the charge angular
distribution with respect to the reconstructed event direction.
These variables have been designed to exploit the different topologies of the
signal and background events, such as short and fuzzy tracks for electrons,
long tracks and sharp rings for muons, two tracks for pions, etc.
For any given PMT, at a distance $r_i$ from the event vertex, the corrected
time $t_{corr}^{(i)}$ is the measured PMT time, $t_i$, corrected for the event
time, $t_0$, and the time of flight:
\[ t_{corr}^{(i)} = t_i - t_0 - \frac{r_i}{c_n}, \]
where $c_n$ is the speed of light in oil, while $\cos\theta_i$ mesures the
cosine of the angle between the PMT location and the event direction.
By binning the sets of $(t_{corr}^{(i)})$ for the hit PMTs and $(\cos\theta_i)$
for all PMTs (including the no-hit ones) and recording the hits, charge, time,
likelihoods information, etc., in each bin, one can construct several hundreds
of potential PID variables.
In addition, some reconstructed physical observables, such as the fitted
Cherenkov-to-scintillation flux ratio, the event track length, the invariant
$\pi^0$ mass, etc. can also serve as PID variables, and they are found to be
quite powerful. 

In principle, all variables which pass the data/MC comparison test can be used
as PID input variables.
However, it is rather well-known that more input variables does not necessarily
imply a better separation in the output, as the PID performance may saturate or
even degrade after the number of input variables exceeds a certain limit that
depends on the problem itself, as well as the PID algorithm.
In the particular case of MiniBooNE, we have found that BDTs can easily handle
large numbers of inputs, whereas ANNs appear to be limited to several tens of
variables~\cite{boo-2}.
Therefore, in order to demonstrate the advantage of the cascade training
technique, we have decided to limit the number of input variables to some
arbitrary small number, e.g., $N_{var}=20$, which can be easily handled by the
ANNs.
However, the 20 variables used by the ANNs may be different than those used by
the BDTs, as we want to utilize a set of inputs that maximizes the performance
of that particular algorithm.
In the following paragraph we briefly discuss the algorithms for chosing
a small subset of input variables from a large pool of input variables.

Boosted decision trees offer several natural ways of ordering the variables
and identifying a subset of $N_{var}$ inputs for maximum performance.
Once the decision trees have been built with all available variables, the
criteria by which the input variables can be easily ordered are, for example:
(a) the order in which they are used,
(b) the frequency with which they are used, or
(c) the number of events they split.
Unfortunately, the neural networks do not allow for such a classification.
However, an alternative, systematic procedure for ordering the input variables
can be defined as follows:
starting from the $i$-th variable, scan all other variables and build a neural
network or a decision tree using only these two inputs, with the $j$-th
variable yielding the best separation.
Using variables $(i,j)$, scan again all other variables and search for a third
variable, say the $k$-th one, which gives a maximum performance for either
a new network or a new decision tree.
The procedure is then repeated to find the 4th, 5th, etc., variable, until the
desired subset size, $N_{var}$, is reached.
Note that starting from different variables may result in a different set of
variables for a given final size $N_{var}$, so an additional loop over the
first variable is needed.

A direct comparison of the different ordering procedures for BDTs is
illustrated in Fig.~1.
The three natural variable ordering schemes for BDTs appear to have similar
performance for $N_{var}>15$, and they appear to be quite close to saturation 
at $N_{var}=50$, the maximum value plotted here.
The systematic ordering appears to have a slightly better performance than
the natural ordering for $N_{var}<20$, in particular when using a relatively
low number of inputs.
The systematic ordering has been carried out rigurously only to $N_{var}=20$
for computational reasons, as the CPU time increases dramatically with
$N_{var}$.
The efficiencies displayed here for $20<N_{var} \leq 25$ have been obtained by
fixing the first $20$ variables and simply searching for {\em additional} new
variables, which may not necessarily yield the best efficiencies.
\begin{figure}[htb]
\begin{center}
\includegraphics[bb = 20 30 550 550, width = 0.87\textwidth]{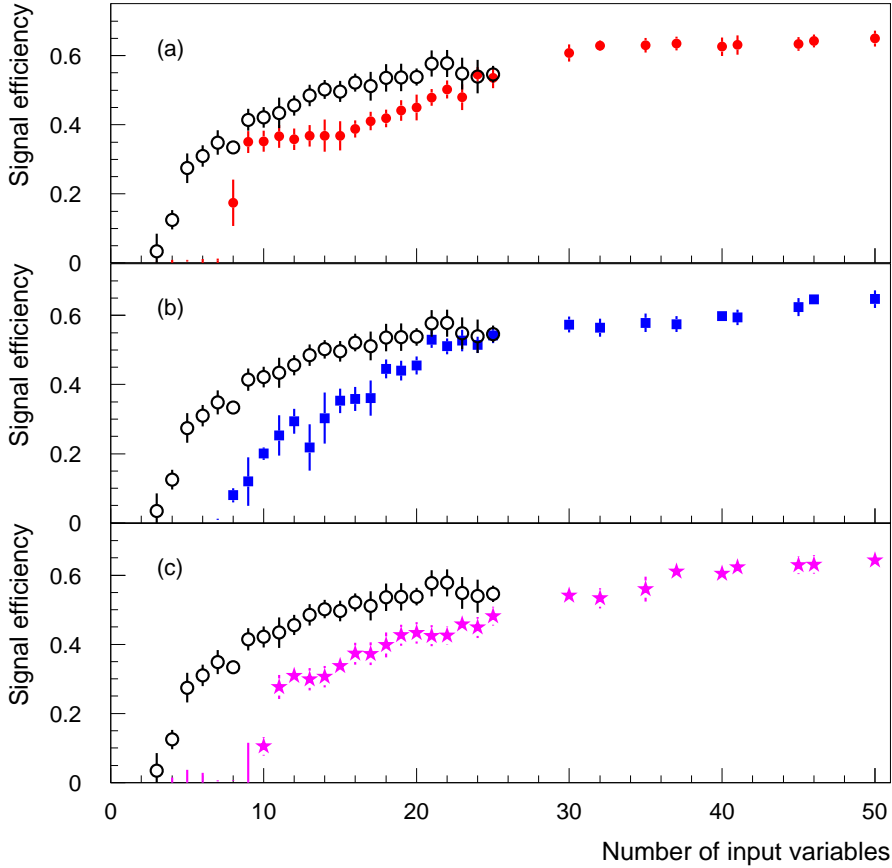}
\end{center}
\vspace{0mm}
\caption{Signal efficiency at 1\% background contamination versus number of
         input variables for boosted decision trees.
         The variable ordering is by (a) usage ordering, (b) usage frequency,
         (c) number of events they split, while the systematic ordering --
         represented by the empty circles -- has been superimposed on all of
         them.}
\label{Fig1}
\end{figure}

For consistency, we apply this systematic procedure for the variable search to
both networks and boosting, using $N_{var}=20$, and turn now to describing the
cascade training. 
%
\section{The Cascade Training Technique}\label{sec:ctt}
%
The idea of the CTT came originally from the combination of the ANN and BDT
outputs~\cite{annboo}.
In general, the ANN output is not 100\% correlated with the BDT output, and
hence one can expect some improvement in the overall performance of the PID
selection algorithm by combining them together. 

Figure~2 below shows the BDT versus the ANN output for the same number
of signal and background events.
For a 1\% background contamination, an ANN cut of about $O_{ANN}>0.98$, or
a BDT cut of about $O_{BDT}>1.15$ is necessary.
By direct observation it is easy to see that the combined OR region
($O_{ANN}>0.98$ or $O_{BDT}>1.15$) helps improve the signal efficiency relative
to a single cut based on either one of the two outputs: the efficiency
yields now $62.5\%$, for a slightly higher contamination level of $1.26\%$.
Insisting on the nominal background contamination level, a signal efficiency
of $54.7\%$ can be reached for $O_{ANN}>0.99$ \underline{or} $O_{BDT}>1.22$, or
even $55.9\%$ for $O_{ANN}>0.92$ \underline{and} $O_{BDT}>1.03$ by optimizing
the cuts.

\begin{figure}[htb]
\begin{center}
\includegraphics[bb = 20 30 550 550, width = 0.9\textwidth]{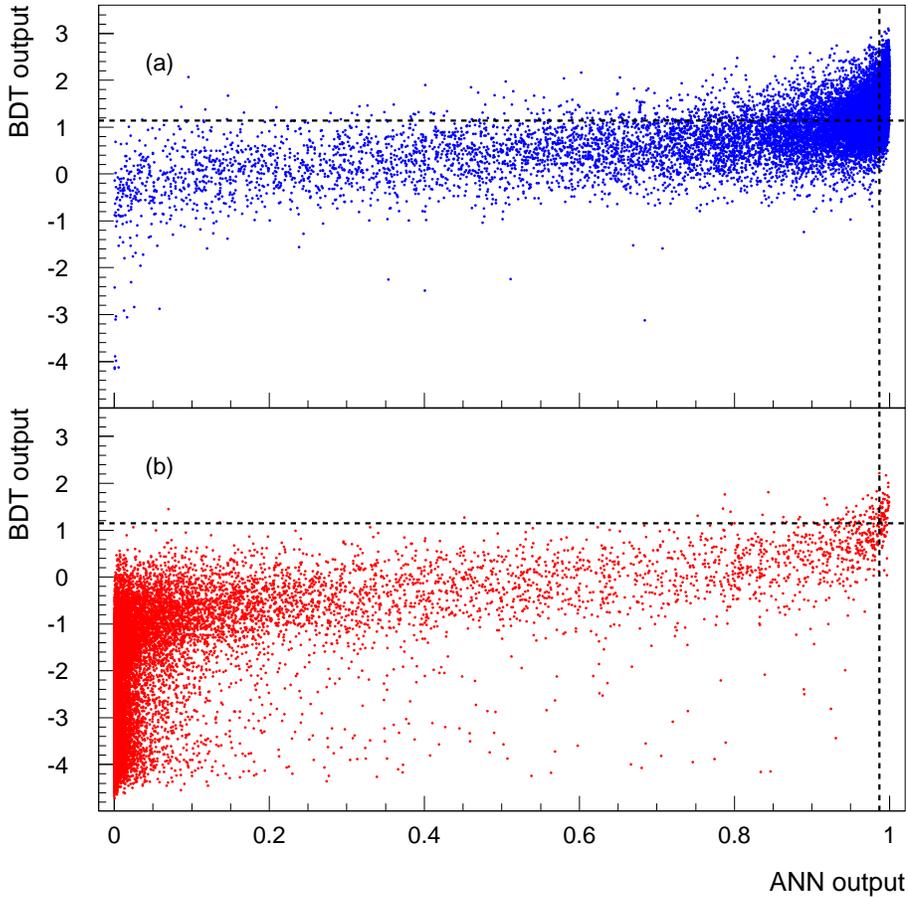}
\end{center}
\vspace{0mm}
\caption{BDT versus ANN output for (a) signal and (b) background events.
         The vertical and horizontal dashed lines indicate the ANN and BDT cuts
         for a 1\% background contamination, respectively.}
\label{Fig2}
\end{figure}

There are three ways to combine the ANN and the BDT outputs in a relatively
straightforward manner. 
The first is to optimize the cuts on both of them -- as discussed in the
previous paragraph.
The second is to take them as inputs to another ANN or BDT.
The third is to use the output of a single algorithm (either ANN or BDT) as
a cut to remove a large portion of the background events, and then force
another algorithm to focus on separating the remaining signal and background 
events -- which are now harder to separate.
We have found that this latter technique proves to be the most efficient one,
especially in the very low background contamination region.  

The CTT procedure can be formulated as follows:
\begin{enumerate}
\item[(a)] Prepare three independent sets of samples {\bf A}, {\bf B}, and
           {\bf C}, where the number of background events in {\bf B} is several
           times larger than that in {\bf A}. 
\\
\item[(b)] Train the PID algorithm (ANN or BDT) with sample {\bf A}, where the
           input variables may be selected as described in
           Section~\ref{sec:vars}, if needed.
\\
\item[(c)] Examine the PID output distribution using sample {\bf C} in order to
           determine the PID cut value, which is set by the point where the
           signal and background distributions cross each other, as shown in
           Fig.~3.
           The crossing point obviously depends on the relative number of
           signal and background events in the sample;
           in this study we have used equal numbers of signal and background
           events, for both the training and test samples.
\\
\item[(d)] Select the training events from sample {\bf B} using the PID cut
           determined by the procedure in step (c) above.
\\
\item[(e)] Train another PID algorithm (ANN or BDT) with the training events
           obtained in step (d) above, where another variable selection
           (as described in Section~\ref{sec:vars}) may be applied.
\\
\item[(f)] Test the performance of the resulting PID algorithm with
           sample {\bf C}. 
\end{enumerate}
Thus, the first ANN or BDT algorithm as built in step (b) only serves to
determine the cut used to select the training events for the second PID
algorithm training in step (e), and hence the name of the cascade training
technique.
%
\section{Results}\label{sec:results}
%
Figure~3 shows the ANN/BDT output distributions from sample {\bf C}
after the first step training of the two algorithms on sample {\bf A}.
In this particular case the signal/background crossing occurs at $O_{ANN}=0.36$
for neural networks and $O_{BDT}=0$ for the boosted decision trees.
Therefore, the events with $O_{ANN}>0.36$ or $O_{BDT}>0$ are selected to form
the restricted sample of events, {\bf B'} and {\bf B''}, respectively, used for
the second step training.
In our case approximately 90\% of the signal events pass either one of these
cuts, while about 90\% of the background events are rejected.
Note that the two peaks seen in the BDT output distribution of background
events in Fig.~3(b) (dashed histogram) represent the two different
bakgrounds in MiniBooNE:
the left-most peak around $O_{BDT}=-4$ corresponds to muon events, which are
relatively easy to identify, while the right peak at about $O_{BDT}=-1$
corresponds to neutral pion events, which are harder to identify.
\begin{figure}[htb]
\begin{center}
\includegraphics[bb = 20 30 550 550, width=0.9\textwidth]{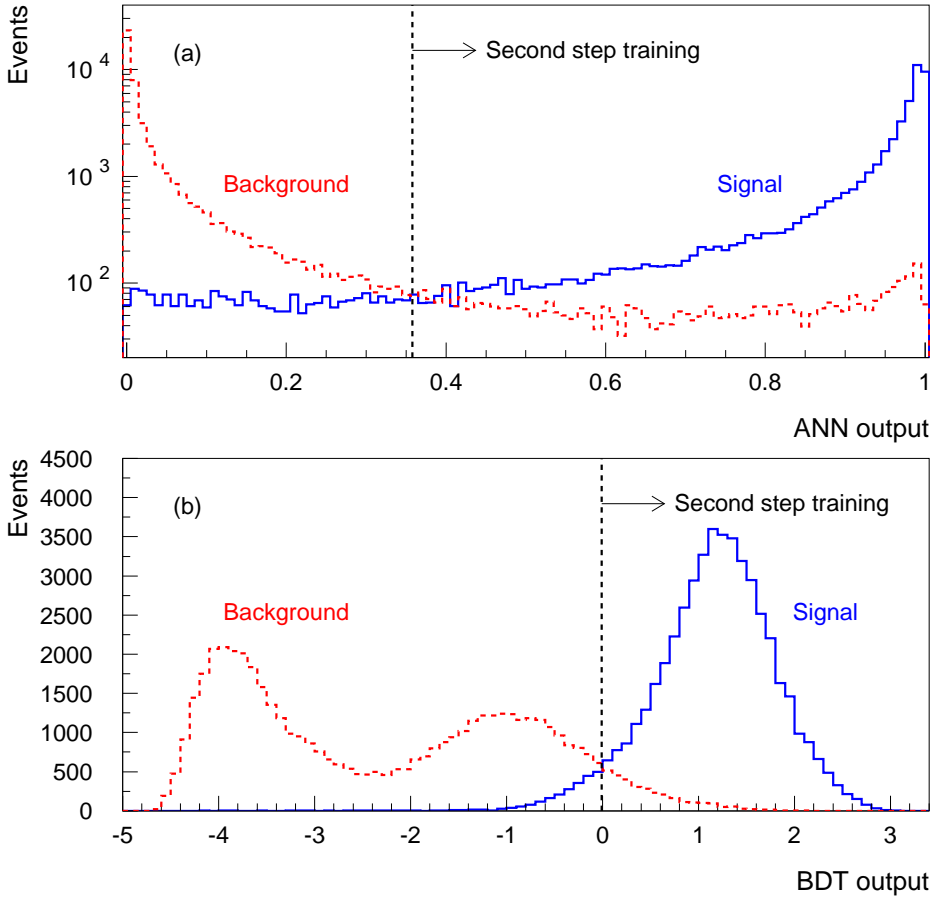}
\end{center}
\vspace{0mm}
\caption{PID output distributions of the test events after the first step
         training for (a) ANNs and (b) BDTs.
         The vertical dashed lines indicate the cuts applied to select the
         training events for the second step training.}
\label{Fig3}
\end{figure}

The efficiencies of the new ANN and BDT algorithms, trained on the new samples,
{\bf B'} and {\bf B''}, respectively, are illustrated in Fig.~4 as
a function of the background contamination level, along with the efficiencies
of the corresponding algorithms trained on the common sample {\bf A}.
The variable sets used in the first and second step training are identical
(although they may be different between ANNs and BDTs).
\begin{figure}[htb]
\begin{center}
\includegraphics[bb = 20 30 550 550, width=0.9\textwidth]{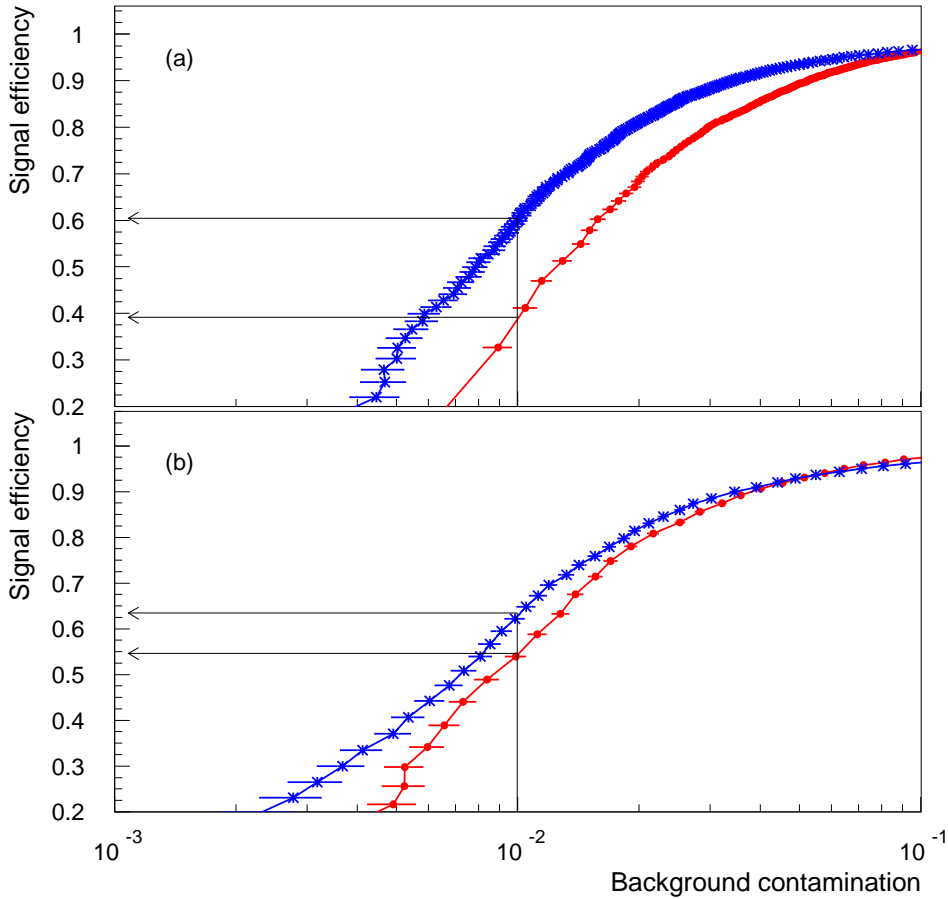}
\end{center}
\vspace{0mm}
\caption{ Signal efficiency versus background contamination for (a) ANNs and
          (b) BDTs for the first and second step training
          (red circles and blue stars, respectively).}
\label{Fig4}
\end{figure}

At 1\% background contamination, the ANN-based signal efficiency increases from
about 39\% to 60\% after the cascade training, which represents a 50\%
improvement.
This improvement is even more significant at lower background contamination
levels, where the signal efficiency can more than double.
For boosting however, the improvement by the cascade training appears to be
less significant, as it raises the signal efficiency from about 54\% to
63\% (an improvement of 17\%) at 1\% background contamination.
As in the ANN case, the relative improvement is also more significant at lower
levels of contamination.

The lower gain in improvement seen when using boosted decision trees may
indicate that the boosting algorithm is already powerful enough to exhaust to
a large extent the different information content between signal and background
events after the first training step.
Nonetheless, while the BDTs may provide a superior PID performance than the
conventional ANNs (54\% versus 39\%), the difference after the CTT is much
reduced (63\% versus 60\%).
This implies that the CTT can significantly help the classification algorithms
exhaust any useful signal-to-background difference, and push the algorithms to
do their best in separating signal from background.  

The efficiencies at the 1\% contamination level are summarized in Table~1.
In addition, the table also gives the efficiencies at a lower contamination
level, $0.5\%$, which shows that the relative gains obtained by using the CTT
are more significant at lower levels of background contamination.
\begin{table}[htb]
\begin{center}
\begin{tabular}{|c|c|c|c|}
\hline\hline
First     & Second    & Efficiency (\%) & Efficiency (\%) \\
algorithm & algorithm & at 0.5\% BCL    & at 1.0\% BCL    \\ \hline\hline
ANN & --  &  8.4 & 38.7 \\
ANN & ANN & 32.5 & 60.3 \\
ANN & BDT & 39.8 & 67.6 \\ \hline
BDT & --  & 25.0 & 54.0 \\
BDT & BDT & 37.0 & 62.8 \\
BDT & ANN & 34.2 & 55.3 \\ \hline\hline
\end{tabular}
\vspace{0mm}
\caption{Signal efficiencies at two different background contamination levels
         (BCL) for the cascade training after initial selection based on ANNs
         (top half), and on BDTs (bottom half).}
\end{center}
\end{table}
%
\section{Conclusions}\label{sec:conclusions}
%
In conclusion, a very efficient way to improve ANN- or BDT-based particle
identification is introduced, namely the cascade training technique.
The procedure is described in detail, as well as the relevant variable 
construction and selection method. 

Our study shows that the CTT can help both ANN and BDT algorithms exhaust the
available information contained in the input variables, while significantly
improving the PID performance in the low background contamination regions.
However, the results reported here are based only on the MiniBooNE detector MC.
The relative improvement obtained by the CTT should depend on the concrete
experiment/application, variable set, and analysis goal.
It could be less, or even more than the numbers presented in this paper.

The intuitive explanation of the high efficiency of CTT may be that if the
algorithm can separate hard-to-identify events, it may be able to identify
easily separable events quite naturally.
Therefore, the strategy is to force the algorithm to focus on learning the
different information content from the very signal-like background events and
true signal events, while disregarding some number of background-like signal
events.
Based on our experience, we believe that in general, in multi-component
background composition case, the CTT should be a very efficient way to improve
the PID efficiency.   

Finally, this technique reveals that in addition to the search of good input
variables and application of a powerful classification algorithm, an apropriate
manipulation of the training event selection opens another way to improve the
efficiency of particle identification.
\section*{Acknowledgements}
%
We are grateful to the entire MiniBooNE Collaboration for their excellent 
work on the Monte Carlo simulations and the software packages for physics
analysis.
It is a great pleasure for Y.\ Liu to dedicate this paper to his Ph.D.
supervisor, Prof.\ Mo-Lin Ge, on the occasion of his seventieth birthday.

This work has been supported by the US-DoE grant numbers
DE-FG02-03ER41261
and
DE-FG02-04ER46112.
%

\end{document}